# Space charge limited photocurrents and transient currents in CdZnTe radiation detectors

Katarina Ridzonova,[*] Eduard Belas, Roman Grill, Jakub Pekarek, and Petr Praus

Charles University, Faculty of Mathematics and Physics, Institute of Physics, Ke Karlovu 5, CZ-121 16, Prague 2, Czech Republic

*We combined steady-state photoconductivity and laser-induced transient current measurements under above-band-gap illumination to study the space charge formation in CdZnTe. Analytical as well as numerical models describing space charge limited photocurrents were developed and an excellent agreement with measured data was obtained especially with the Drift-diffusion model. Linear rise of photocurrent at low bias was observed and ascribed to the trapping of injected holes at the region close to the cathode side. Influence of space charge formation, photoconductive gain, contribution of shallow and deep levels to photocurrent-voltage characteristics were numerically simulated. According to the measurements and calculations, recent principles used at the evaluation of detector properties, mainly the mobility-lifetime product, via the photoconductivity are critically assessed.*

## I. INTRODUCTION

The influence of space charge on the charge transport, space-charge-limited currents (SCLC), and the shape of currents-voltage characteristics in semiconductors equipped with injecting contacts are extensively studied for decades starting with the pioneer works of Lampert [1] and Mott & Gurney [2]. Overview of recent examinations may be found for example in [3]. Materials are mostly investigated through the effect of space charge build up and respective screening of the electric field with the single-carrier formula neglecting diffusion current

$$J_{SCLC} = \frac{9}{8}\frac{\epsilon_0 \epsilon_r \mu_e \theta V^2}{L^3}, \qquad (1)$$

where $\varepsilon_0$, $\varepsilon_r$, $\mu_e$, $V$ and $L$ are vacuum and relative permittivity, electron mobility, bias and thickness of planar sample, respectively. The electron injection from the cathode dominating in the circuit is assumed. The material is characterized by the ratio of the free electron density $n$ and the sum of the free and trapped electron density $n_t$

$$\theta = \frac{n}{n + n_t}, \qquad (2)$$

where the free-to-trapped electron density ratio is calculated in the nondegenerate Boltzmann statistics in the form independent of the Fermi energy

$$\frac{n_t}{n} = \frac{gN_t}{N_C}e^{\frac{E_t}{k_B T}} \qquad (3)$$

and $N_C$, $g$, $N_t$, $E_t$, and $T$ are effective density of states in the conduction band, degeneracy factor of the trap, trap density, depth of trap relatively to the conduction band, and temperature, respectively. Space charge stored in the trap usually dominates above the free electron one and $\theta=n/n_t$ is routinely used. Properties of dominant trap responsible for the carrier trapping and the space charge formation are deduced by the Arrhenius-type fit combined with other techniques like photo-induced current transient spectroscopy, thermally stimulated current measurements, e.g. [4], [5].

---

[*] katka.ridzonova@gmail.com

The exploration of materials and devices by the space-charge limited photocurrents (SCLP) offers significantly wider set of experiments and representative data characterizing investigated system than one could extract from SCLC. Using the above-band-gap light absorbed in the close proximity to the illuminated semi-transparent electrical contact allows researchers to maintain a tuneable source of carriers as a substitution of carriers supplied by the injection contact in SCLC. The space charge may be thus created regardless of the type of the contact. In addition, the variable excitation may be also used at the investigation of the dynamics of the space charge formation with much better precision than it could be arranged by the bias pulsing. Other option is connected with the investigation of the material and contact homogeneity by scanning selected contact regions. In contrast to evident advantages, parallel creation of both electrons and holes and their trapping near the contact may affect their electrical properties, which could manifest in the variation of the dark current component. Effects related with this entanglement should be considered at the analysis to eliminate possible faults at the data evaluation.

Measurement of photocurrent-voltage characteristics (PV) at above-bandgap illumination has become a common method to evaluate important transport properties such as the carrier mobility-lifetime product $\mu\tau$. The mostly used approach to fit the dependence of photocurrent $J$ on internal electric field $E$ is based on equation (4) attributed to Many [6], which in the case of above-bandgap light illuminating cathode reads:

$$J = \frac{J_0}{1+\frac{s_e}{\mu_e E}} \frac{\mu_e \tau_e E}{L} \left[1 - \exp\left(-\frac{L}{\mu_e \tau_e E}\right)\right]. \tag{4}$$

Here $J_0$ represents saturation photocurrent, which is proportional to the excitation intensity, and $L$ is the thickness of planar sample. The electron transport characteristics are described by mobility $\mu_e$, lifetime $\tau_e$, and surface recombination velocity $s_e$. Considering strong attenuation of the above-bandgap light absorbed typically at the depth of less than 1 μm near the surface, photogenerated holes are almost immediately drained by the cathode under applied bias while electrons drift through the bulk to the anode. Thus only electrons contribute to the overall photoconductivity and the problem is limited to the solution of single-particle-type task. Analogously, the hole signal can be recorded by simply reversing the applied bias polarity or by irradiating the opposite contact. Other presumptions, in which equation (4) is valid, are suppressed carrier injection from surface states and spatially constant $\mu_e$ and $\tau_e$. In case of electrical contacts directly deposited on the sample's surface ohmic contacts and flat bands at the metal-semiconductor interface should be maintained. It should be mentioned that the original derivation of eq. (4) did not assume the steady state bias. Pulsed field and blocking electrodes detached from contacts by insulating spacer were used. The key assumption for the validity of equation (4) is the spatially uniform internal electric field $E=V/L$ for applied voltage $V$, which equals to the clause of zero space charge distributed in the sample, either induced by non-ohmic contacts or by the photo-excitation.

Many´s equation (4) was routinely used at the determination of $\mu\tau$ in various semiconductor materials such as CdZnTe [7–10], HgI$_2$ [11], [12], Tl$_4$CdI$_6$ [13], SbSeI [14], TlSi$_4$ [15], Cs$_2$Hg$_3$S$_4$ [16], hexagonal boron nitride [17–20], halide perovskite CH$_3$NH$_3$PbI$_3$ [21], CsPbBr$_3$ films [22] and others. However, in all cases it is hard to deduce, if the key assumption of uniform electric field was fulfilled, since the measurement of the PV characteristics does not reveal the presence of space charge in the sample. If this requisite was not met, another model involving the influence of space charge would have to be used [9,23,24] or an experiment should be set in order to maintain a homogeneous electric field. Space charge can be reduced or fully prevented by using a chopped regime of illumination instead of continuous one, as it was done in [10,14–16,22,25]. Even in such experimental conditions, however, equation (4) should be used cautiously as the space charge relaxation could proceed with much slower dynamics than its formation and a memory effect could debase the results as well.

Laser-induced transient current technique (L-TCT) represents nowadays the perspective tool, how to non-destructively investigate the interior of semi-insulating materials with extended lifetime of at least one of photo-generated carriers [26]. The detail analysis of the current waveform shape enables us to evaluate the drift mobility of carriers and their lifetime as well as the electric field and the space charge profile along the sample thickness [27]. It may be thus used as independent technique at the verification of theoretical predictions offered by the theory of SCLP.

In this article, we report on the theoretical and experimental investigation of SCLP. We have developed three photoconductivity models under above-bandgap illumination which involve space charge formation in addition to surface recombination. We also propose a method based on the laser-induced transient current technique which allows us to clearly prove or refute the presence of space charge within the studied material. According to the result one can decide whether the measured PV characteristics might be evaluated by Many´s equation (4) or by models involving space charge. All measurements were performed on CdZnTe with 15% zinc content, which is a suitable semiconductor material for fabrication of room temperature X-ray and gamma-ray detectors. It is distinguished by relatively high atomic number, wide energy band gap (~1.5 eV) and high resistivity [28].

## II. THEORY

### A. Space charge limited photocurrents 1

In this section we introduce the simplest analytical model describing SCLP, further denoted SCLP1. It is advantageous for the initial characterization of the most important processes occurring in sample subjected to the above-band-gap cathode illumination and ensuing space charge formation. The model considers one shallow trap and neglects the dark currents and photo-generated holes.

The shallow trap allows us to use the non-degenerate Boltzmann statistics defining its occupancy. By neglecting the concentration of free electrons in the dark $n_0$ and trapped electrons $n_{t0}$ relatively to the photo-excited electrons, the space charge density $\rho$ may be expressed as

$$\rho = -e(n - n_0 + n_t - n_{t0}) \cong -e\frac{n}{\theta} \qquad (5)$$

where $\theta$ is defined in eq. (2) and $e$ is the elementary charge. Combining eq. (5), Poisson´s equation and Ohm´s law in the same way as it is done at the derivation of SCLC form (1) [29] one may express the spatial profile of the electric field $E(x)$

$$E(x) = -\sqrt{E^2(0) + \frac{2jx}{\epsilon_0 \epsilon_r \mu_e \theta}}. \qquad (6)$$

Here $j$ represents the current density and $I_l$ is the incident light intensity.

The boundary condition represented by the electric field just below the illuminated contact $E(x=0)$ in equation (6) may be obtained by solving the balance equation describing the surface dynamic equilibrium between the optical excitation, surface recombination and drift into the bulk

$$I_l - s_e n(0) - \frac{j}{e} = 0, \qquad (7)$$

which yields in

$$E(0) = -\frac{s_e j}{\mu_e (I_l e - j)}. \qquad (8)$$

The term connected to the direct or deep-level-mediated interband recombination in the sample's bulk was omitted in eq. (7) due to negligible dark currents (since almost none free holes are present in the semiconductor volume, electrons cannot recombine there and we can assume infinite carrier lifetime $\tau_e$). Integrating equation (6) through the sample thickness $L$ we may express the applied bias $V$ dependent on $j$

$$V = \frac{\epsilon_0 \epsilon_r \mu_e \theta}{3j} \left\{ \left[ \left( \frac{s_e j}{\mu_e (I_l e - j)} \right)^2 + \frac{2jL}{\epsilon_0 \epsilon_r \mu_e \theta} \right]^{3/2} - \left( \frac{s_e j}{\mu_e (I_l e - j)} \right)^3 \right\}. \qquad (9)$$

A formula similar to equation (6) was also derived by Chatenier [23]. When the drift term in eq. (7) is much less than other two ones, eq. (9) may be converted to eq. (1). Let us note that equation (9) involves the steady state current and doesn't contain any parameter defining interaction of free electrons with the trap. Consequently, this model cannot be used for determination of quantities defining electron (de)trapping such as carrier capture cross-section, lifetime as well as the mobility-lifetime product. Comparing these findings with Many's equation (4) it is evident that the application of eq. (4) to the steady-state experiment described by eq. (9) in the effort to determine µτ is unsupported.

The SCPL1 may be easily generalized also to a model of several shallow levels, if their occupancy is ruled by the Boltzmann statistics. In such a case the parameter θ may be obtained as

$$\theta = \frac{n}{n + \sum_i n_{ti}}, \qquad (10)$$

where the summation runs over respective level occupancies $n_{ti}$ and other formulas remain unchanged.

*1. Transient currents in case of inhomogeneous space charge distribution*

The laser-induced transient current technique (L-TCT) is a powerful tool for investigation of principal electronic properties of semi-insulating materials determining the electric field profile in semi-insulating semiconductors by analysing the current waveform (CWF) shape [26,27,30–33]. It is common in CdZnTe radiation detectors, that the space charge is induced by blocking contacts ensuring low dark current and carrier depletion. This setup leads in a formation of nearly constant space charge in the whole sample, unless fully screened inactive region appears [23]. The CWF has an exponential dependence on time $I(t) \sim e^{-ct}$ where the damping factor $c$ is related to the space charge and mobility-lifetime product $\mu_e \tau_e$ [26].

Conversely, for inhomogeneous space charge density induced by injecting contacts or by above-bandgap illumination we have to deal with equations (6), (8) defining the internal electric field. According to Ramo theorem [34] the CWF induced by charge $Q$ moving with velocity $v$ through a planar sample of thickness $L$ can be generally expressed as [26]

$$i(t) = \frac{Q(t)v(t)}{L}. \qquad (11)$$

The trajectory of electron excited at the cathode is easily calculated by integrating the kinetic equation $dx/dt = \mu_e E(x)$, in which $E(x)$ defined by equation (6) is used

$$x(t) = \frac{\epsilon_0 \epsilon_r \mu_e \theta}{2j} \left[ \left( \frac{jt}{\epsilon_0 \epsilon_r \mu_e \theta} - \frac{s_e j}{\mu_e (I_l e - j)} \right)^2 - \left( \frac{s_e j}{\mu_e (I_l e - j)} \right)^2 \right]. \qquad (12)$$

Time derivation and substituting into (11) finally leads to CWF in the form

$$i(t) = \frac{Q(t) \mu_e}{L} \left( \frac{s_e j}{\mu_e (I_l e - j)} + \frac{jt}{\epsilon_0 \epsilon_r \theta} \right). \qquad (13)$$

In case of a good quality sample, when the carrier lifetime significantly exceeds the time of the passing of carriers through the detector (called transit time), time independent $Q(t)=Q_0$ may be used and $i(t)$ attains a linear shape. As results from equation (13), increasing transient current confirms the presence of a negative space charge in the detector while decreasing current shape after the onset corresponds to a positive space charge.

### B. Space charge limited photocurrents 2

In the second model called SCLP2 we generalize the previous model SCLP1. In addition to SCLP1, we assume the general energy of deep trap levels possibly located near the Fermi energy at which the Fermi-Dirac statistics defining level occupancy instead of Boltzmann statistics must be used. In addition, we also consider substantial density of free electrons supplied by ohmic cathode. The integrating the Gauss´ law we obtain the following transcendental equation for internal electric field:

$$-A[E(x) - E(0)] + B\ln\left|\frac{E(x) - E_1}{E(0) - E_1}\right| + C\ln\left|\frac{E(x) - E_2}{E(0) - E_2}\right| = \frac{e}{\epsilon_0 \epsilon_r} ADx. \tag{14}$$

Detailed derivation of the equation (14), corresponding to the SCLP2 model, and the meaning of individual parameters *A, B, C, D, E₁, E₂,* and *E(0)* can be found in the Appendix. The electric field profile *E(x)* is calculated numerically for each $x \in (0, L)$ and fixed trap parameters solving Eq. (14) with given *j*, obtained from the experiment. Respective bias *V* is acquired by the *E(x)* integration. Whole PV dependency is obtained by repeating the procedure with different *j*. Similarly as in the case of SCLP1, we see that SCLP2 does not involve any parameter defining (de)trapping or the carrier lifetime. Electrons may leave the sample only at contacts; by the surface recombination at the cathode and by drain at the anode.

### C. Drift-diffusion model

Analytical models SCLP1 and SCLP2 are based on the significant simplification, which could make them useless in general conditions and obtained results inaccurate. There is especially neglected contribution of holes, which may affect the transport characteristics in the high resistivity n-type samples and even more markedly in p-types. The involvement of holes into the calculations disables utilization of analytical approach and fully numerical treatment must be engaged. We adopted for this purpose previous codes for the simulation of transient currents solving in parallel drift-diffusion and Poisson's equations in material with traps [35]. In addition, we refined the boundary condition defining the interface carrier density by involving an interface layer which supresses the surface recombination. Instead of fixed free carriers density directly defined by the metal-semiconductor Schottky barrier we use a dynamical prescription according [36],[37] and define the free carrier current at the contacts in the form

$$J_e(0) = -\gamma_e^{(0)} e \left[n_0^{(0)} - n(0)\right], \tag{15a}$$

$$J_h(0) = \gamma_h^{(0)} e \left[p_0^{(0)} - p(0)\right], \tag{15b}$$

$$J_e(L) = \gamma_e^{(L)} e \left[n_0^{(L)} - n(L)\right], \tag{15c}$$

$$J_h(L) = -\gamma_e^{(0)} e \left[p_0^{(L)} - p(L)\right], \tag{15d}$$

where $J_{e(h)}(0)$, $n_0(p_0)^{(0)}$, $J_{e(h)}(L)$, and $n_0(p_0)^{(L)}$ are the equilibrium electron (hole) current density and dark density of electrons and holes at the cathode ($x=0$) and anode ($x=L$), respectively. Respective γ-factors define the transfer rate through the interface. In addition to [36],[37], where only limiting cases γ=0 and γ=∞ were considered, we use a general γ at the fit of experimental data. Let us note that $n_0(p_0)$ may be different at the cathode and the anode due to possibly unlike band banding at the interfaces.

The improvement of the contact model is indispensable for the right description of the current in biased sample and the surface recombination at the illumination. While the case with γ=0 suppresses the surface recombination by the drain of respective carrier to the contact, it completely disables electric current. Conversely, γ=∞ induces large surface recombination of carriers excited by the above-band-gap light in the thin layer <1μm below the semitransparent contact. We apply the contact model defining the interface transfer of carriers by the linear functions of the carrier density (15a)-(15d) for description of basic properties of the current-carrying illuminated interface with general value of the γ-factor.

Drift-diffusion electron and hole equations involving carriers (de)trapping on defect levels described by the Shockley-Read-Hall model were numerically time-integrated [35]. Adaptive Step-size Control of the fourth-order Runge-Kutta method was used for integration. Electrostatic potential and the electric field profile were calculated by solving the Poisson's equation at each integration step. The simulation of the steady state and chopped photoconductivity occurring in the period significantly exceeding the drift time or lifetime of free carriers is done with electrons and holes in a steady state distribution. This simplification does not notably disturb the obtained results but it significantly speeds up the calculation and allows us making simulations in the reasonable short time period without putting big demands to the computational resources [35].

### III. EXPERIMENT

All measurements were performed on n-type semi-insulating planar CdZnTe (CZT) detector which was grown by High Pressure Bridgman Growth Method. The sample, further marked with the letter A, was mechanically polished by SiC abrasive and chemically etched in 3% bromine-methanol solution for one minute, after which it was rinsed in methanol and isopropyl alcohol. Gold contacts were deposited from 1% $AuCl_3$ aqueous solution onto both largest sides of the detector with dimensions 6 mm × 4 mm × 1.4 mm. In addition, seven other samples not shown in this paper were measured revealing similar behaviour as observed on presented sample.

Electron mobility-lifetime product of detector A determined by the alpha spectroscopy was $1.2 \times 10^{-3}$ $cm^2 \cdot V^{-1}$ and its resistivity was $7.6 \times 10^{10}$ Ω·cm.

PV characteristics under continuous illumination were measured by laser diode powered by Tektronix AFG 3252 arbitrary waveform generator. The wavelength of incident light was 690 nm (hν ≈ 1.87 eV) and intensity of 4 μW/$cm^2$. Homogeneously irradiated area of 3 $mm^2$ was used. Current through the detector was measured indirectly on 1 MΩ load resistor. Since we irradiated the cathode side of the samples with above-bandgap light only electrons contributed to the measured signal. Final photocurrent was determined as a difference between current under illumination and dark current.

Laser-induced transient currents were excited from the cathode side by probing laser pulses with the 2 ns width, 100 Hz repetition frequency and above-band-gap 662 nm wavelength. The laser diode was powered by a Picosecond pulse generator. Acquired CWFs were amplified by high-frequency bipolar 3-GHz Miteq AM-1607-3000R amplifier with signal conversion of 6.85 mV/$\mu$A into a voltage pulse and then were recorded using a 4-GHz digital LeCroy oscilloscope [38]. The laser pulse intensity and repetition frequency were tested to be low enough so that the photocurrent as well as the current waveform shape were not influenced by the laser pulses.

### IV. RESULTS AND DISCUSSION

## A. Photocurrent

In order to compare theoretical models we measured PV characteristics under continuous regime of cathode illumination. Experimental data for sample *A* together with fits by Many´s equation, SCLP1 and Drift-diffusion models are shown in Fig. 1. We can see that measured photocurrent vs bias curve may be split respectively to the regions of nearly linear rise up to the voltage ~200 V and saturation above 200 V. In the linear region, the space charge induced by photo-electrons captured in bulk trap levels significantly screens the bias near cathode and only part of photo-electrons may pass to the anode while a lot of them disappear near cathode due to the surface recombination. At the photocurrent saturation the screening of electric field near the cathode is suppressed and most of the photo-electrons are collected. The maximum photocurrent at the saturation is obviously limited by the intensity of illumination $j=eI_l$. In some samples, the saturation part of the photocurrent can be tilted due to contact effects, mainly by the photoconductive gain caused by a large space charge formed by trapped photo-holes near the cathode [39]. We observed such effect in several our samples. The chosen sample *A* does not show such effect and the saturation is well defined.

Comparing the Many's fit with SCLP1 model, we may clearly observe a much better agreement of SCLP1 with the experiment. Similar disaccord of the Many's fit with experiment may be identified also in works [8], [9]. Fit parameters obtained from Many´s equation are $s_e/\mu_e$ =140 Vcm$^{-1}$ and $\mu_e\tau_e$ = 2.3 × 10$^{-4}$ cm$^2$V$^{-1}$, which is five times lower value than the $\mu_e\tau_e$ obtained from alpha spectroscopy.

The fit by SCLP1 model under assumption of fixed mobility $\mu_e$ = 1057 cm$^2$V$^{-1}$s$^{-1}$ for CdZnTe [35] gives the surface recombination velocity s = 170 000 cms$^{-1}$ and $\theta$ = 5 × 10$^{-5}$. SCLP2 model considering experimentally determined $n_0$ = 7.8 × 10$^4$ cm$^{-3}$ leads to almost identical fit as SCLP1, since the best results for sample *A* are obtained in presence of shallow trap with $E_t$ < 0.54eV at which both models coincide.

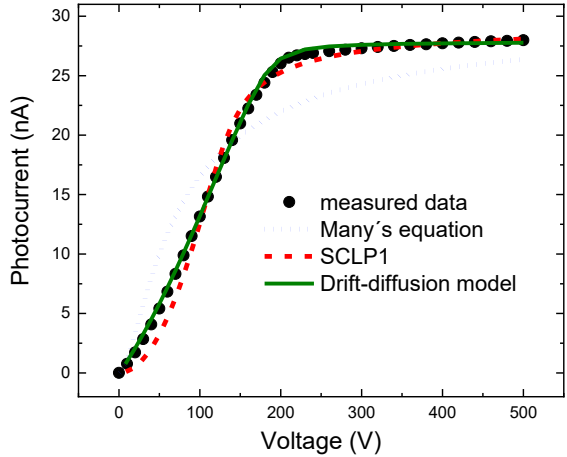

FIG. 1. PV characteristic of sample *A* gained under continuous laser diode illumination with the wavelength 690 nm, measured in the configuration for collecting electrons only. Measured data are fitted according to the Many´s equation (dotted line), SCLP1 model (dashed line) and Drift-diffusion model (green line), respectively.

For the explicit illustration of the effect of various parameters to the photocurrent, we plot in Fig. 2 the SCLP1 fit together with curves with changed parameters $\theta$, $I_l$, and $s_e$, respectively. Initial parameters corresponding to black curve were $\theta$ = 5 × 10$^{-5}$, s = 70 000 cm/s and $I_l$ = 2 × 10$^{12}$ cm$^{-2}$s$^{-1}$. In case of reduced $\theta$ the larger screening of the electric field induced by respective deep level implies slower onset of the PV characteristics and delayed saturation. Reduced $I_l$ is represented by correspondingly reduced saturation current. Increased $s_e$ makes the transfer between increasing and saturated parts of the PV characteristics more gradual and the curve is similar to the Many´s equation model. It is seen that the parabolic shape of the PV characteristics at low bias persist at all situations.

The comparison of models SCLP1 and SCLP2 is shown in Fig. 3, where the degenerate Fermi-Dirac statistics was enforced to the SCLP2 model setting $n_1 = n_0$. The parameters of SCLP1 model remained the same as for black curve in Fig. 2, while parameters for SCLP2 model were $\theta = 7 \times 10^{-7}$, $s = 70\,000$ cms$^{-1}$, $I_l = 2 \times 10^{12}$ cm$^{-2}$s$^{-1}$, $n_1 = n_0 = 7.8 \times 10^4$ cm$^{-3}$. It is seen that the degenerate level occupancy results even in slower photocurrent rise which corresponds to larger polynomial exponent than the 2$^{nd}$ power resulting in SCLP1. Such slow photocurrent rise was observed in work [7]. We deduce that from the onset of the photocurrent it is possible to roughly estimate whether shallow or deep levels prevail in the material. The accuracy of the estimation may be, however, limited by participation of holes in space charge formation.

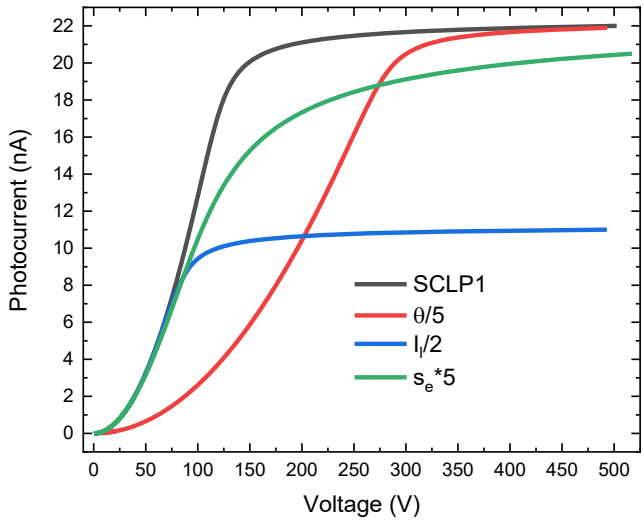 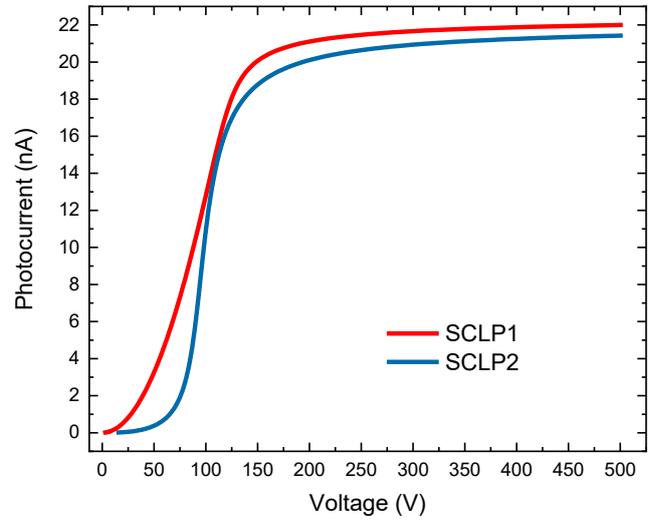

FIG. 2.
Influence of change of $\theta$ (red line), intensity (blue line) and surface recombination (green line) on the shape of PV characteristics (black line) calculated by SCLP1 model.

FIG. 3.
Comparison of SCLP1 model with defect level below Fermi energy (red line) and SCLP2 model with defect level on Fermi energy (blue line).

Although the SCLP1 and SCLP2 models fit the experiment in Fig. 1 pretty well, we see distinct deviations mainly in the region of low biasing at the incomplete carrier collection manifesting in the convex bending of SCLP curves. The measured photocurrent has a linear behaviour rather than the parabolic shape gained from the SCLP models. These deviations were interpreted as a consequence of simplifications used at the derivation of the SCLP models. Based on this consideration we engaged the Drift-diffusion model described in Section 2.3 and searched for a convenient defect and contact properties, which describe experimental data and which are consistent with the mobility-lifetime product determined from the alpha-spectroscopy.

In Fig. 4 we can see the measurement of PV characteristics at versatile excitation intensities along with an almost identical fit in which three defect levels and flat bands at metal-semiconductor interface were assumed. Levels are characterized by concentration $N_t$, energy $E_t$ and by capture cross-sections of electrons and holes $\sigma_e$ and $\sigma_h$ listed in the Table I. The energy levels 1 and 2 are important at the fit of the photoconductivity while the level 3 was added to fit the appropriate lifetime evaluated from the alpha-spectroscopy. The Fermi energy ($E_F$) of the semi-insulating sample was set to the midgap region 0.76 eV below the conduction band. Chosen $E_F$ is consistent with measured resistivity of the material. Material parameters for $Cd_{0.9}Zn_{0.1}Te$ are the bandgap energy $E_g$= 1.57 eV [28] and the hole mobility $\mu_h$ = 77 $cm^2Vs^{-1}$ [40]. The absorption coefficient corresponding to the photon energy 1.87 eV was deduced according [41] as $\alpha$ = 25 000 $cm^{-1}$. The transfer rate defining the transfer of electrons through the cathode $\gamma_e^{(0)} = 5 \times 10^4$ cm $s^{-1}$ was obtained by the fit. Other transfer rates $\gamma_h^{(0)}, \gamma_e^{(L)}, \gamma_e^{(0)}$ did not influence the fit markedly and they were not fitted. They were fixed at a high value $10^7$ cm $s^{-1}$, which affords the surface density of respective free carriers defined directly by the Schottky barrier.

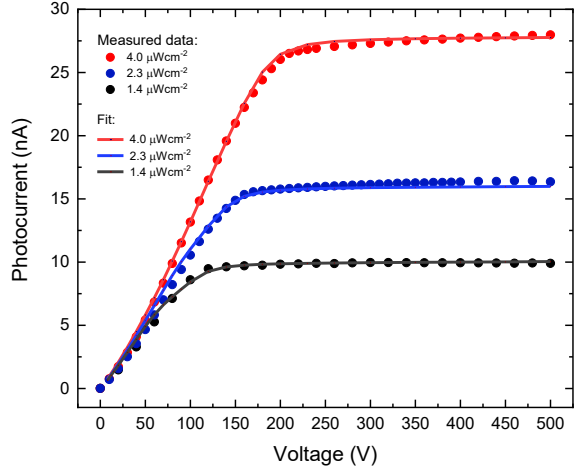

FIG. 4. PV characteristics measured on sample *A* at different light intensities (4 µW/cm², 2.3 µW/cm² and 1.4 µW/cm²) with the wavelength 690 nm. Measured data are fitted according to the Drift-diffusion model with parameters listed in Table I.

Important characteristics of the fit are outlined in Fig. 5, where the electric field and space charge density (Fig. 5a), level occupancy (Fig. 5b) and free carrier density (Fig. 5c) profiles calculated at the 30 V biasing are shown. Photo-excited electrons are primarily trapped on the level 1 forming negative space charge in the bulk of the sample. As a consequence, the electric field is screened near the cathode side and is increasing towards the anode, which is depicted as a red line in Fig. 5a.

The important feature distinct from the predictions of SCLP models is the shoulder clearly seen at the position ≈ 0.05 cm at all profiles. The reason of this artefact lies in the injection of holes from the anode and their participation at the space charge formation. While the hole density is low at the region $x \in (0.5, 1.4)\ mm$ due to higher electric field, it increases nearby cathode for $x < 0.5\ mm$ due to the lowered electric field. Consequently, the enhanced hole trapping close cathode, mainly on the level 2, yields in a significant compensation of the negative space charge formed by the photo-electrons in the sample. The width of this compensated region is dependent on applied bias and enlarges with the decreasing bias. Since the deep electron trap level 1 is significantly filled by trapped electrons, the free electron lifetime increases and the electrons may pass the compensated region without strong trapping or recombination. The photocurrent then exceeds the parabolic-like theoretical prediction of the SCLP models and attains nearly linear shape at the low bias. The photocurrent linear growth at the low bias observed in other works [8–10,17–20,42] may be also assigned to the above-described phenomena in which injected holes from anode participate at the charge transport. In case of non-saturated photocurrent the interpretation of the PV characteristics by strong surface recombination should be considered as well.

The positive space charge formed in the close vicinity of the cathode, see Figs. 5a, b, is caused by the trapping of photo-excited holes on the level 2. This effect results in the enhancement of the electric field near the cathode and increases injection of electrons from the contact. In the investigated sample $A$ the injecting electrons are significantly damped by the low transfer rate $\gamma_e^{(0)}$. Larger $\gamma_e^{(0)}$ would lead to the emphasized photoconductive gain [39] and no saturated photocurrent could be seen.

Though the number of parameters defining the fit looks pretty large, a simplification of the model and reduction of the number of parameters, for example by joining levels 1 and 2 to an appropriate single level was not successful. It is also worth to mention that similar model with two midgap levels was successfully applied at the explanation of depolarization of CdZnTe radiation detector by the above-band-gap light [43]. Summarizing, the presented model with two levels positioned near the Fermi energy and another shallower electron trap conveniently depicts important features of the charge dynamics in the sample and consistently complies with all experimental results.

| Energy level | $N_t$ (cm$^{-3}$) | $E_t$ (eV) | $\sigma_e$ (cm$^2$) | $\sigma_h$ (cm$^2$) |
|---|---|---|---|---|
| 1 | $3 \times 10^{11}$ | 0.66 | $2 \times 10^{-14}$ | $1.0 \times 10^{-14}$ |
| 2 | $9 \times 10^{11}$ | 0.85 | $3 \times 10^{-16}$ | $2.5 \times 10^{-15}$ |
| 3 | $2 \times 10^{12}$ | 0.50 | $1 \times 10^{-14}$ | 0 |

TABLE I. Fitting parameters of defect levels using the Drift-diffusion model to describe experimental data plotted in Figs 1, 4 and 6.

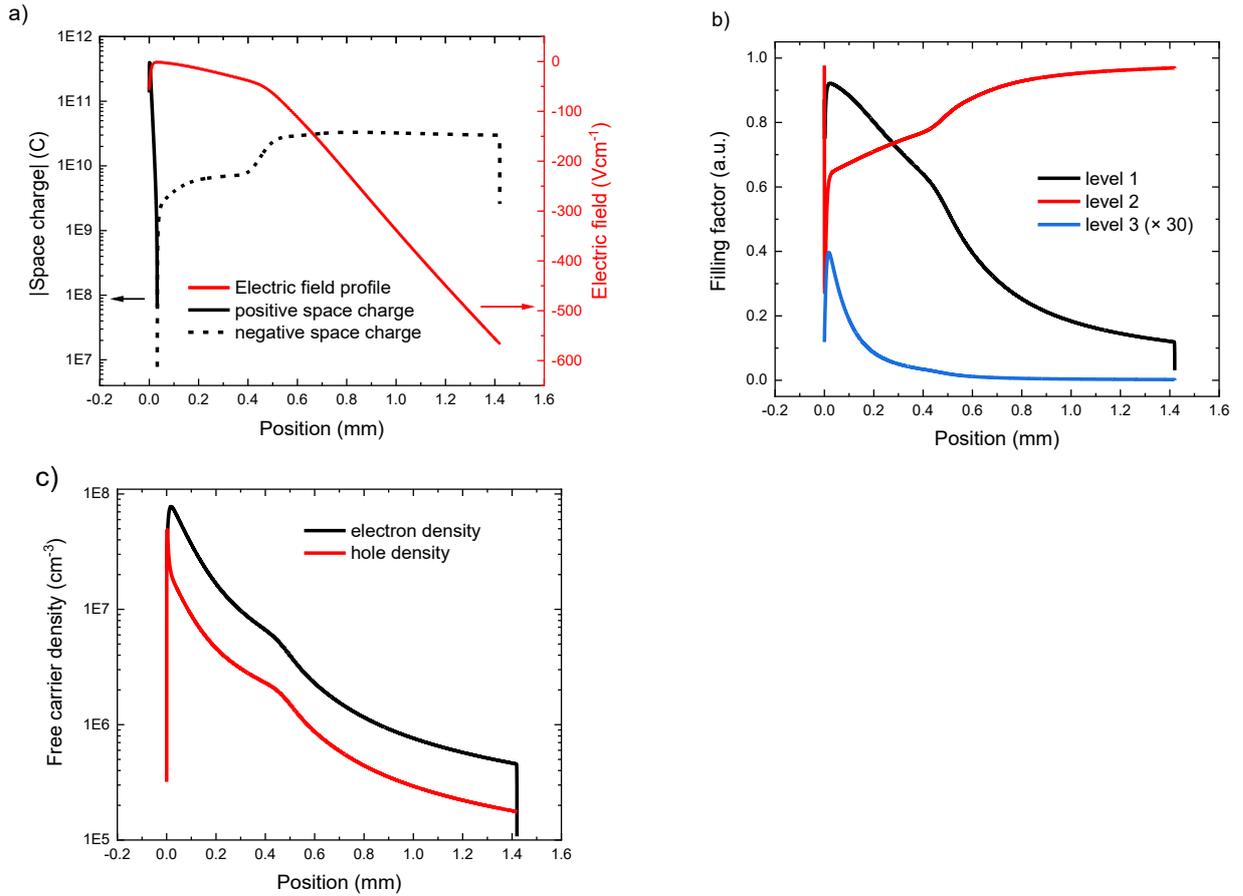

FIG. 5. (a) Space charge and electric field profile, (b) defect level occupancy (blue line was multiplied by factor 30) and (c) free electron and hole density calculated by Drift-diffusion model for bias 30 V and illumination intensity 2.3 μWcm$^{-2}$ with fit parameters listed in Table I.

## B. Transient current measurement

To prove the existence of space charge in CZT under continuous above-bandgap illumination we further applied the L-TCT on sample *A* in two regimes.

First we illuminated the cathode of the sample by probing laser pulses only. Laser intensity and frequency were low enough so that the illumination did not influence the properties of the sample but served only for the characterization of internal electric field. The space charge originating from trapped photo-generated carriers was insignificant in this case. Examples of CWFs collected at different bias 30 V, 100 V, and 300 V are plotted in Figs. 6(a)-6(c), respectively, by red curves labelled 'without LED'. The descending CWF apparent at all biases is dominantly induced by the finite electron lifetime $\tau_e$=1.2μs derived from the alpha-spectroscopy, which is fitted by parameters defining level 3. The descending CWF descending at 30 V has the damping factor *c* higher than it could be explained only by determined $\mu\tau$, which proves the weak positive charging of biased sample. The positive space charge appearing in this setup could be created due to the variation of carrier density induced by slightly blocking cathode or injecting anode gold contacts. Such effect is hourly observed in Au/n-CdZnTe/Au radiation detectors[30]. Weak undulations at the waveforms are partly due to sample inhomogeneity and also caused by the adjoint electronics[30].

In the second regime, the cathode was simultaneously illuminated by the above-described probing pulses and by continuous LED diode with the wavelength of 690 nm. In this case we simulated the conditions, which occur in the sample during continuous photocurrent measurements. Since the L-TCT records only deviations from the steady state, the generated greater amount of charge carriers in the case of additional continuous illumination should not contribute to the total current pulse. The variations of CWFs are caused by the warping of electric field induced by the negative space charge appearing due to the photo-carrier trapping. The principal effect of the cathode illumination on the transient current profile is clearly demonstrated in Figs. 6(a)-6(c) by waveforms labelled 'LED 690 nm'. The significantly suppressed electric field at low biasing results in a strong surface recombination and reduced collected charge as depicted in Fig. 6(a). By increasing bias, the electric field near the cathode grows in parallel as the screening fades away, surface recombination attenuates and the waveforms attain their undisturbed shape as it is apparent in Fig. 6(c). The crossing between screened and unscreened cathode is ideally illustrated by the CWF in Fig. 6(b) where the waveform though significantly tilted reveals features similar to the regular shape. Ascending CWF is consistent with the negative space charge created by trapped photo-generated electrons under continuous illumination, which induces the ascending transient current, in which the CWF damping due to lifetime is overcome. Moreover, the charge collection efficiency at continuous illumination and high biasing results were even higher than those obtained in the dark. We interpret this effect as the result of the filling of electron traps by electrons created by steady excitation.

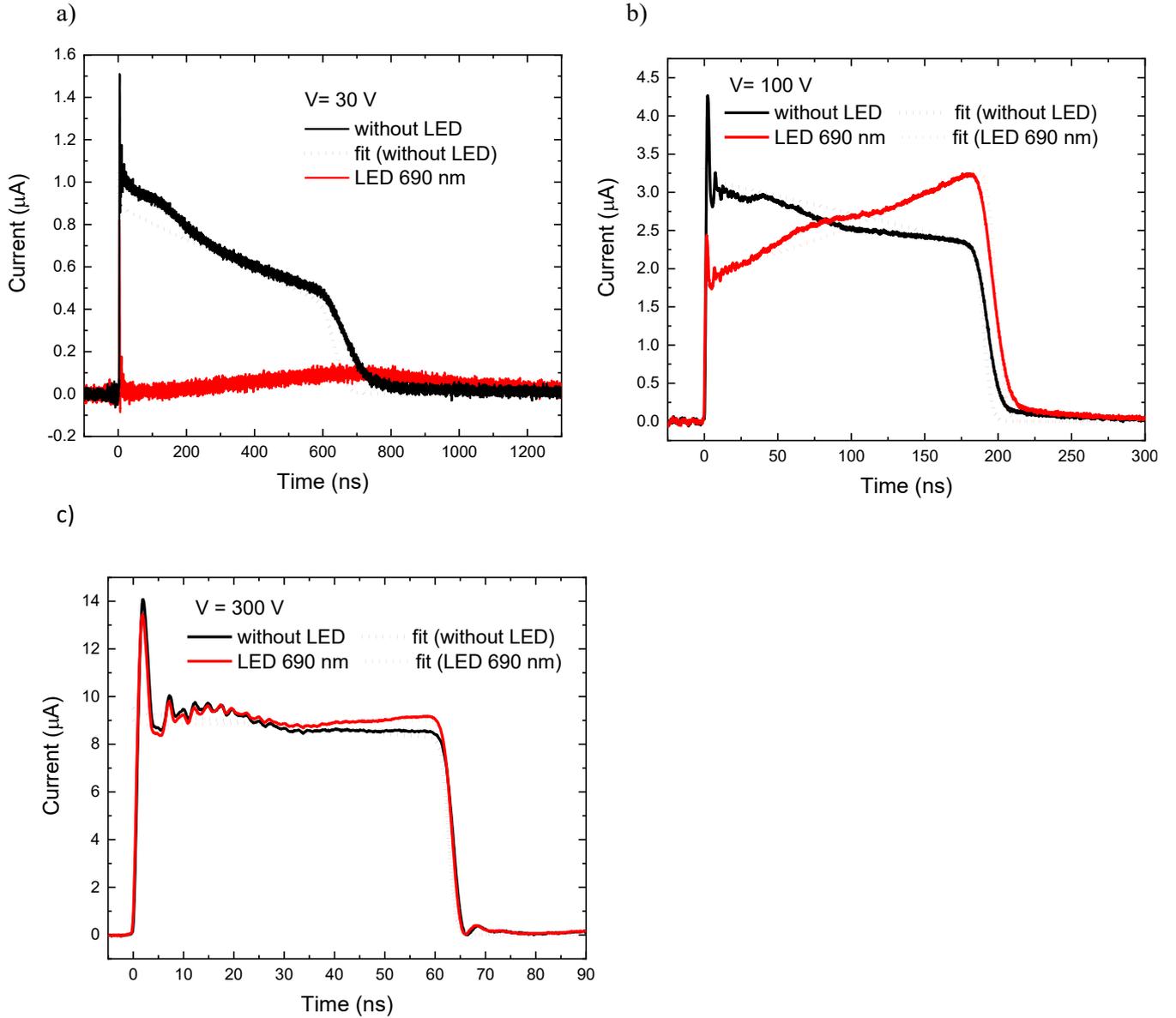

FIG. 6. (a)- (c) Current waveforms measured on sample *A* by L-TCT using laser pulses only (red line) and with additional continuous LED illumination 690 nm (black line) for bias a) 30 V, b) 100 V and c) 300 V. Dotted lines represent the fit using Drift-diffusion model with parameters listed in Table I.

### C. Many's equation

Photocurrent-voltage characteristics are routinely used at the evaluation of the mobility-lifetime product and surface recombination considering the Many's equation (4). It is seen in Fig. 1 that the fit with eq. (4) neither approximate the experiment satisfactorily nor launches out $\mu_e\tau_e$ evaluated with the alpha spectroscopy. The reason primarily consists in the experimental conditions, when the space charge formation significantly violated the homogeneous electric field in the sample presumed at the eq. (4) derivation. With aim to assess this eventuality we performed the simulation with the Drift-diffusion model using the defect structure given in Table I and taking a very low excitation intensity $I = 2.5\times10^9$ photons cm$^{-2}$s$^{-1}$, which is 1000× less than the maximum intensity used in Fig. 4. The photo-induced space charge formation was then significantly suppressed. Such measurement of the photoconductivity is limited only to samples with low noise so that a good quality signal could be retrieved. The calculations were done in the dynamic regime of the photoconductivity measurements. Both bias and light were switched on at t=0 on sample in the equilibrium and the current was obtained after respective period $t_d$ of steady illumination and biasing. Dark current was calculated by the same way just without switching on the light. This approach ensures the well-defined setting of the sample evolution without inevitable memory effects, which should be anticipated in real chopped experiment. The photocurrent was obtained obviously by the differentiation of the values. The results are shown in Fig. 7 where the PV characteristics for different $t_d$ are plotted. We also add the relevant graphs of Many's function (4) with the right $\mu_e\tau_e$=1.2×10$^{-3}$ cm$^2$/V (dotted line) and another $\mu_e\tau_e$=5.2×10$^{-3}$ cm$^2$/V (dashed line).

We may see that exact $\mu_e\tau_e$ may be obtained in the short-pulse setup when $t_d$ < 40 μs. This finding is consistent with the L-TCT measured at the pulsed bias [27], where exact $\mu_e\tau_e$ unaffected by the polarization may be determined.

Extending $t_d$ to about 1 ms shows continuous deviation from the right curve to significantly larger $\mu_e\tau_e$=5.2×10$^{-3}$ cm$^2$/V. The reason of this shift consists in the detrapping properties of the level 3, see table 1, acting as the electron trap. The detrapping time of this level is 1 ms. Using $t_d$ similar or larger than the detrapping time allows trapped carriers to come into equilibrium with free electrons in the conduction band and the level becomes invisible at the time-resolved experiments. Omitting level 3, the remaining levels 1 and 2 define $\mu_e\tau_e$=5.2×10$^{-3}$ cm$^2$/V and the comparison of calculated profiles with the respective (dashed) curve perfectly agrees with this expectation

The further increase of $t_d$ to $t_d$ > 0.1 s shows continuous deflection of the curve especially at low bias, which could be interpreted as further improvement of $\mu_e\tau_e$. Such interpretation is, however, incorrect. Right reason consists in (i) the charging of deep levels due to the electron depletion induced by blocking cathode defined by low $\gamma_e^{(0)}$ and (ii) by the trapping of photo-holes near the cathode. While the item (i) induces the positive space charge within the bulk of the sample, the process (ii) entails in the formation of a large positive space charge in a thin layer in the cathode. The activity of both processes leads to an enhancement of the electric field near the cathode and the increase of the photocurrent at low bias - photoconductive gain.

To conclude, the validity of Many´s equation is in practice limited mainly by three factors. The first is a negative space charge from trapped photo-electrons which leads to inhomogeneous electric field profile. The second factor corresponds to the case of experiment in which delay time $t_d$ is similar or larger than the detrapping time of shallow level. In this case shallow level is not reflected in the PV measurement. As it has been already mentioned in section A, Level 3 doesn´t even strongly influence the fitted shape of PV curves in Fig. 4, where only energy levels 1 and 2 are important at the fit of the photoconductivity. Level 3 was here added according to the knowledge of the lifetime from alpha spectroscopy. The third frequent phenomena limiting relevance of Many´s equation is a positive space charge in the vicinity of cathode which may lead to photoconductive gain.

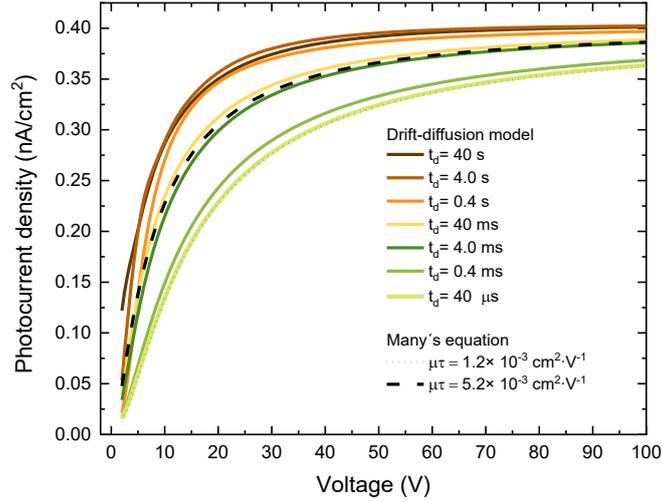

Fig. 7. Drift-diffusion model of PV characteristics obtained at chopped regime of illumination with different delay time $t_d$ (full lines) and their comparison with Many´s equation (dashed and dotted line) with various mobility life-time products.

## V. CONCLUSION

Steady-state photocurrents were measured on planar CdZnTe radiation detectors under above-bandgap illumination and were theoretically studied by analytical and numerical models. Three models, SCLP1, SCLP2 and Drift-diffusion model, describing space-charge limited photocurrents were developed and conveniently applied in the fit of experimental data with much better precision then it could be obtained within a previous approach using Many´s formula. Additionally, we have created a model describing transient currents in case of inhomogeneous space charge distribution induced by cathode illumination. Within SCLP1/SCLP2 models we showed that degeneracy level leads to slower photocurrent onset corresponding to larger polynomial exponent than the 2$^{nd}$ power resulting in SCLP1. We conclude that shape of the onset can be used as a rough estimation whether shallow or deep levels dominate in the material.

Numerical Drift-diffusion model explained consistently all experimental data, the PV characteristics at various illumination intensities and the transient current measured at different biasing in the sample held both in the dark and with the illuminated cathode. Three defect levels with energies 0.66 eV, 0.85 eV and 0.50 eV were ascribed to the measured sample. Experimentally observed linear onset of the photocurrent at low biases which is also frequently appearing in multiple works was explained by compensation of the negative space charge nearby the illuminated cathode by trapped injected holes.

Numerical simulation of PV characteristics at low intensities and chopped regime showed that validity of Many´s equation may be in practice limited not only by negative space charge from trapped photo-electrons. Phenomena of photoconductive gain caused by trapping of photo-generated holes in the vicinity of cathode also reduces usability of Many´s equation. Incorrect determination of mobility-lifetime product may also be due to the fact that shallow levels with detrapping times lower than period of illumination cannot be detected in PV characteristics.

The collected achievements allow us to conclude that the measurements of steady-state photo-current, if they are used as a single analytical method, do not enable researchers to evaluate the carrier lifetime as well as other carrier dynamic parameters. The measurement of PV characteristics should be completed by the independent measurement of mobility-lifetime product such as transient current technique or alpha spectroscopy measurements. Measurement of PV characteristics at various intensities together with knowledge of carrier lifetime then give detailed image about defect structure of material using Drift-diffusion model. Certain shallow levels may not be distinguished in defect structure just from PV measurement without considering lifetime obtained from other experiment.

## ACKNOWLEDGMENTS

This project was financially supported by the Grant Agency of the Czech Republic under Grant No. P102/ 18-12449S, by the Grant Agency of Charles University under Project No. 1374218 and by Charles University Project No. SVV–2019–260445.

## APPENDIX

In the SCLP2 model, Fermi-Dirac statistics must be used to describe occupancy on deep trap level. Thus, the concentration of trapped electrons may be expressed as $n_t = \frac{N_t}{1+n_1/n}$ with the parameter $n_1 = N_c \, exp[E_t/(kT)]$ being related to the trap level position. Considering also the dark concentration of free electrons $n_0$ and trapped electrons $n_{t0}$, the Gauss' law may be used in the form

$$\frac{dE}{dx} = -\frac{e}{\epsilon_0 \epsilon_r} \left[ -\frac{j}{e\mu_e E} + \frac{N_t}{1 - \frac{n_1 e \mu_e E}{j}} - n_0 - \frac{N_t n_0}{n_0 + n_1} \right]. \tag{A1}$$

By converting square bracket in eq. (A1) to a common denominator we get the second order polynomial in E(x) in the numerator. If we mark $E_1$ and $E_2$ the roots of the polynomial, equation (A1) then attains the form:

$$\frac{E\left(1 - \frac{n_1 e \mu_e}{j} E\right)}{(E - E_1)(E - E_2)} dE = -\frac{1}{\epsilon_0 \epsilon_r \mu (n_0 + n_1)} dx. \tag{A2}$$

Corresponding roots $E_1$, $E_2$ are

$$E_{1,2} = \frac{n_1 + N_t - D \pm \sqrt{(D - n_1 - N_t)^2 + 4n_1 D}}{-2AD}, \tag{A3}$$

where meaning of parameters $A$ and $D$ is given in eq. (A4).

Numerical SCLP2 model given by the eq. (14) is obtained by applying partial fraction decomposition to the integrand on the left side of eq. (A2) and then by doing partial integration. All parameters $E_1, E_2, A, B, C$ and $D$, appearing in the equation (14), are listed in eq. (A3) and in following eq. (A4)

$$A = \frac{n_1 e \mu_e}{j}, B = \frac{E_1(1 - AE_1)}{E_1 - E_2}, C = \frac{E_2(1 - AE_2)}{E_2 - E_1}, D = n_0 + \frac{N_t n_0}{n_0 + n_1}. \tag{A4}$$

The boundary condition for SCLP2 model

$$E(0) = \frac{e(I_l + n_0 s_e) - j - \sqrt{4ejn_0 s_e + (eI_l - j + en_0 s_e)^2}}{2e\mu_e n_0} \tag{A5}$$

was attained from the balance equation involving dark current and electron density:

$$I_l - s_e(n - n_0) - \frac{j - j_d}{e} = 0. \tag{A6}$$

The generalization to the multiple level design cannot be easily introduced here since the roots of respective polynomials appearing in equation (A2) cannot be calculated analytically. Nevertheless, the direct numerical integration of equation (A1) does not make big troubles as well.

CdZnTe radiation detectors by above-bandgap light. *J. Appl. Phys.*, **117**, 165702 (2015).